\documentclass[aip,jcp,reprint]{revtex4-1}

\usepackage{color}
\usepackage{dcolumn}
\usepackage{amsmath}
\usepackage{amssymb}
\usepackage{mathtools}
\usepackage{graphicx}
\usepackage{latexsym}
\usepackage{amsfonts}
\usepackage{amssymb}
\usepackage{comment}
\DeclareGraphicsExtensions{.pdf,.gif,.jpg}

\newcommand{\be}{\begin{equation}}
\newcommand{\ee}{\end{equation}}

\def\a{\alpha}
\def\b{\beta}
\def\g{\gamma}

\def\eps{\epsilon}
\def\w{{\omega}}
\def\im{{\mathrm{i}}}
\def\ex{{\mathrm{e}}}
\def\ud{\mathrm{d}}
\def\hc{\text{h.c.}}

\def\bunitmat{\mbox{$\mathbf{1}$}}
\def\bH{\mbox{\boldmath $h$}}
\def\bG{\mbox{\boldmath $G$}}
\def\bS{\mbox{\boldmath $\varSigma$}}

\begin{document}


\title{Electronic transport in molecular junctions: The generalized Kadanoff--Baym ansatz with initial contact and correlations} 



\author{Riku Tuovinen}
\affiliation{QTF Centre of Excellence, Turku Centre for Quantum Physics, Department of Physics and Astronomy, University of Turku, 20014 Turku, Finland \looseness=-1}

\author{Robert van Leeuwen}
\affiliation{Department of Physics, Nanoscience Center, University of Jyv{\"a}skyl{\"a}, 40014 Jyv{\"a}skyl{\"a}, Finland \looseness=-1}

\author{Enrico Perfetto}
\affiliation{Dipartimento di Fisica, Universit{\`a} di Roma Tor Vergata, Via della Ricerca Scientifica 1, 00133 Rome, Italy \looseness=-1}
\affiliation{INFN, Sezione di Roma Tor Vergata, Via della Ricerca Scientifica 1, 00133 Rome, Italy \looseness=-1}

\author{Gianluca Stefanucci}
\affiliation{Dipartimento di Fisica, Universit{\`a} di Roma Tor Vergata, Via della Ricerca Scientifica 1, 00133 Rome, Italy \looseness=-1}
\affiliation{INFN, Sezione di Roma Tor Vergata, Via della Ricerca Scientifica 1, 00133 Rome, Italy \looseness=-1}


\begin{abstract}
The generalized Kadanoff--Baym ansatz (GKBA) offers a computationally inexpensive approach to simulate out-of-equilibrium quantum systems within the framework of nonequilibrium Green's functions. For finite systems the limitation of neglecting initial correlations in the conventional GKBA approach has recently been overcome [Phys. Rev. B \textbf{98}, 115148 (2018)]. However, in the context of quantum transport the contacted nature of the initial state, i.e., a junction connected to bulk leads, requires a further extension of the GKBA approach. In this work, we lay down a GKBA scheme which includes initial correlations in a partition-free setting.
In practice, this means that the equilibration of the initially correlated \emph{and} contacted molecular junction can be separated from the real-time evolution. The information about the contacted initial state is included in the out-of-equilibrium calculation via explicit evaluation of the memory integral for the embedding self-energy, which can be performed 
without affecting the computational scaling with the simulation time and system size.
We demonstrate the developed method in carbon-based molecular junctions, where we study the role of electron correlations in transient current signatures.
\end{abstract}

\maketitle 


\section{Introduction}\label{sec:intro}

State-of-the-art electronic components are engineered from nanoscale building blocks with emerging quantum phenomena~\cite{Keimer2017, Tokura2017, Basov2017, Ronzani2018, Kokkoniemi2020}. These devices are not isolated but affected by a wide variety of environmental conditions and external perturbations such as temperature variations, structural defects, and chemical contamination on the samples. The device operation is typically ultrafast; there is no guarantee for an instant relaxation to a static configuration once the device is switched on. Emerging transient effects depend on, e.g., quantum dynamics and correlations~\cite{Wijewardane2005, Verdozzi2006, Myohanen2008, Uimonen2011, Myohanen2012, Latini2014, Talarico2019, Talarico2020, Tuovinen2020}, system geometry and topology~\cite{Khosravi2009, Vieira2009, Perfetto2010, Rocha2015, Wang2015, Tuovinen2019Nanoscale, Ridley2019, Tuovinen2019NJP}, and the response to external perturbations or thermal gradients~\cite{Croy2009, Kurth2010, Foieri2010, Ness2011, Arrachea2012, Eich2016, Tuovinen2016PNGF, Tuovinen2016PRB, Covito2018, Karlsson2018, Honeychurch2019, Kershaw2020, Bostroem2020}. Recently, pump-probe spectroscopic methods have grown in number rapidly leading to the current field of ultrafast materials science with sub-picosecond temporal resolution being routinely achieved~\cite{Prechtel2012, Cocker2013, Hunter2015, Rashidi2016, Cocker2016, Jelic2017, Chavez2019, McIver2020}.

To address all this, a fully time-dependent quantum description including many-body correlations is necessary, as the individual components of the systems are operating on ultrafast time scales at the quantum level. The nonequilibrium Green's function (NEGF) approach~\cite{Baym1961, kbbook, Keldysh1965, Danielewicz1984, svlbook, Balzer2013book} is a natural choice: The dynamical information about the system, e.g. electric currents or the photoemission spectrum, is encoded into the NEGF. Accessing this information requires solving the equations of motion for the NEGF which is computationally expensive. However, this can be made computationally more tractable by reducing the two-time-nature of the NEGF into a single-time-description. This approach, the generalized Kadanoff--Baym ansatz (GKBA)~\cite{Lipavsky1986}, is a well-established procedure, and it has been succesfully applied in, e.g. molecular junctions~\cite{Galperin2008, Ness2011, Latini2014, Hopjan2018, Karlsson2018, Cosco2020}, and spectroscopical set\-ups for atomic~\cite{Balzer2012, Perfetto2015, Covito2018, Covito2018PRA}, molecular~\cite{Bostroem2018, Perfetto2018jpclett, Perfetto2019jctc, Perfetto2020}, and condensed matter systems~\cite{Perfetto2016, Sangalli2018, Schueler2019, Murakami2020, Tuovinen2020, Schueler2020}.

In the language of the Keldysh formalism~\cite{Keldysh1965,Danielewicz1984,svlbook}, the GKBA approach concerns with real-time Green's functions, namely the lesser and greater Keldysh component. A drawback in this approach is that the so-called mixed Keldysh components, with one of the time arguments imaginary and the other one real, are not included. The role of these mixed components is to relate the equilibrium (Matsubara) calculation to the out-of-equilibrium one, and therefore a consistent description of the initial correlations is troublesome. It has been shown to be possible to include the initial correlations in the GKBA approach as a separate calculation~\cite{Semkat2003, Karlsson2018, Hopjan2019, Bonitz2019} although it has been customary to use a noncorrelated initial state and build up correlations via a time evolution
excluding external perturbations~\cite{Rios2011, Hermanns2012}.

For transport setups also the initial contacting of the molecular junction contributes to the initial correlations. In the \emph{partitioned} approach~\cite{Caroli1971a,Caroli1971b} the initial state is uncontacted, and the molecular region is suddenly brought into contact with the leads. In this case, the initial correlation collision integral due to the contact or embedding self-energy vanishes. The contacted initial state can be constructed by a sudden or adiabatic switching (AS) of the contacts and evolving the system without external fields to a contacted equilibrium. In the \emph{partition-free} approach~\cite{Cini1980} the initial state is contacted, and there is a unique thermo-chemical equilibrium. The information about this coupled equilibrium is then encoded in the initial correlation (IC) collision integral $\mathcal{I}^{\text{ic}}(t)$. In this paper, we derive an expression for $\mathcal{I}^{\text{ic}}(t)$ in closed form for the embedding self-energy. This calculation can be separated from the time-evolution, similar to Ref.~\onlinecite{Karlsson2018}. The derived expression can directly be combined with many-body self-energies, resulting in a partition-free approach to the GKBA time-evolution for an initially correlated and contacted transport setup.

The paper is organized as follows. In Sec.~\ref{sec:model} we introduce the model system and the governing equations of the GKBA approach (with the underlying NEGF theory detailed in Appendix~\ref{app:details}). We outline the calculation of the initial contacting collision integral in Sec.~\ref{sec:ic} and defer the implementation details to Appendix~\ref{app:embedding} and~\ref{app:contour}. Then, in Sec.~\ref{sec:results} we present numerical simulations for time-resolved electronic transport in carbon-based molecular junctions. We draw our conclusions and discuss future prospects in Sec.~\ref{sec:concl}.

\section{Model and method}\label{sec:model}

\begin{figure}[t]
\centering
\includegraphics[width=0.475\textwidth]{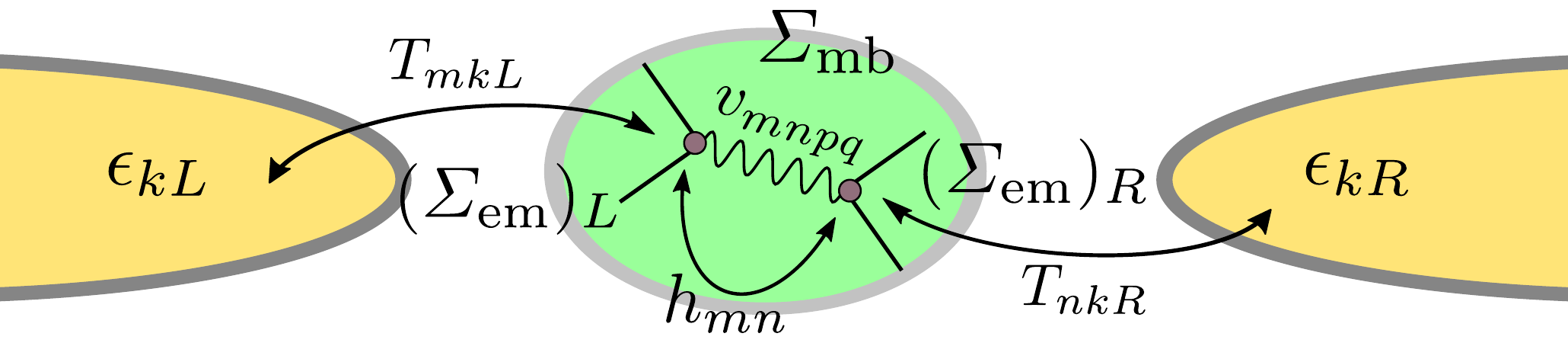}
\caption{Schematic molecular junction described by Eq.~\eqref{eq:hamiltonian}. The molecular structure is represented by the single-particle matrix elements $h_{mn}$ and the interaction vertex $v_{mnpq}$, which is taken into account by the many-body self-energies $\varSigma_{\text{mb}}$. The molecular device is connected to $\a$ leads (only two depicted, $\a\in\{L,R\}$) via the tunneling matrix elements $T_{mk\a}$, represented by the embedding self-energies $(\varSigma_{\text{em}})_\a$.}
\label{fig:setup}
\end{figure}

We consider an electronic junction consisting of a quantum-correlated molecular device ($C$) which is connected to an arbitrary number of noninteracting, metallic leads ($\a$), see Fig.~\ref{fig:setup}. The molecular junction is described in terms of the second-quantized Hamiltonian
\begin{align}\label{eq:hamiltonian}
\hat{H} & = \sum_{k\a,\sigma}\eps_{k\a}\hat{c}_{k\a,\sigma}^\dagger \hat{c}_{k\a,\sigma} + \sum_{mn,\sigma} h_{mn}\hat{c}_{m,\sigma}^\dagger \hat{c}_{n,\sigma} \nonumber \\
& + \sum_{mk\a,\sigma} [T_{mk\a}\hat{c}_{m,\sigma}^\dagger \hat{c}_{k\a,\sigma} + \hc] \nonumber \\
& + \frac{1}{2}\sum_{\stackrel{mnpq}{\sigma\sigma'}}v_{mnpq}\hat{c}_{m,\sigma}^\dagger\hat{c}_{n,\sigma'}^\dagger\hat{c}_{p,\sigma'}\hat{c}_{q,\sigma} ,
\end{align}
where $m,n,p,q$ label a complete set of single-electron states in the molecular device, and $\a$ labels the leads. Above, $\eps_{k\a}$ describes the single-electron energy state $k$ in the $\a$-th lead, $h_{mn}$ are the single-particle matrix elements for the molecular region, $T_{mk\a}$ are the tunneling matrix elements between the molecular device and the leads, and $v_{mnpq}$ are the two-electron Coulomb integrals for the molecular device. The annihilation (creation) operator $\hat{c}_{x,\sigma}^{(\dagger)}$ removes (creates) an electron from (to) state $x$ with spin orientation $\sigma\in\{\uparrow,\downarrow\}$, and they obey the fermionic anti-commutation rules $\{\hat{c}_{x,\sigma},\hat{c}_{y,\sigma'}^\dagger\}=\delta_{xy}\delta_{\sigma\sigma'}$ for indices $x,y$ belonging either to the leads or to the molecular device.

We note that all objects introduced in Eq.~\eqref{eq:hamiltonian} are diagonal in spin space. However, the following consideration could straightforwardly be extended to cases with, e.g., spin--orbit or Zeeman terms in the molecular Hamiltonian~\cite{Tuovinen2019NJP}, or ferromagnetic leads~\cite{Krawiec2006}. In addition, it would be possible to include a contribution from a thermomechanical field to the lead energy dispersion giving rise to a relative temperature shift in the leads~\cite{Luttinger1964, Eich2014, Eich2016, Covito2018JCTC, Perfetto2018}.

To access time-dependent nonequilibrium quantities for the system described by Eq.~\eqref{eq:hamiltonian}, we 
consider the equation of motion
for the single-particle density matrix $\rho$ (see Appendix~\ref{app:details} for background)
\be\label{eq:eom-rho}
\frac{\ud}{\ud t} \rho(t) + \im [h_{\text{HF}}(t),\rho(t)] = - [\mathcal{I}(t) + \mathcal{I}^{\text{ic}}(t) + \hc] ,
\ee
where $h_{\text{HF}}$ is the single-particle Hamiltonian supplemented with the time-local Hartree--Fock (HF) self-energy. The time-nonlocal self-energies due to many-particle and embedding effects appear in the collision integrals
\begin{align}
\mathcal{I}(t) & = \int_{t_0}^t \ud \bar{t} \left[\varSigma^>(t,\bar{t})G^<(\bar{t},t) - \varSigma^<(t,\bar{t})G^>(\bar{t},t)\right] , \label{eq:collint-t} \\
\mathcal{I}^{\text{ic}}(t) & = -\im \int_0^\b \ud \tau \varSigma^\rceil(t,\tau) G^\lceil (\tau,t) , \label{eq:collint-ic}
\end{align}
where $t_0$ marks the time when the system is driven out of equilibrium, and $\b$ is the (equilibrium) inverse temperature. We refer to Appendix~\ref{app:details} for the description of the different components, e.g., the greater ($>$) and lesser ($<$) functions. Solving Eq.~\eqref{eq:eom-rho} is computationally demanding due to the full two-time history of the functions in Eqs.~\eqref{eq:collint-t} and~\eqref{eq:collint-ic}, and because of the self-energies' functional dependency on the Green's functions $\varSigma[G]$.

To reduce the computational complexity the Green's functions are commonly approximated by the GKBA
\be\label{eq:gkba}
G^\lessgtr(t,t') = \mp G^{\text{R}}(t,t') \rho^\lessgtr(t') \pm \rho^\lessgtr(t) G^{\text{A}}(t,t')
\ee
with $\rho^< \equiv \rho$ and $\rho^> \equiv 1-\rho$, and the propagators are described for the coupled system at the HF level
\be\label{eq:propagator}
G^{\text{R}/\text{A}}(t,t') = \mp \im \theta[\pm(t-t')]\mathcal{T}\ex^{-\im \int_{t'}^t \ud \bar{t} [h_{\text{HF}}(\bar{t})\mp \im \varGamma/2]} ,
\ee
where $\mathcal{T}$ is the chronological time-ordering and $\varGamma$ is the tunneling probability matrix from the leads to the molecular region. Here we have used the wide-band approximation (WBA) for the retarded/advanced embedding self-energy, see Appendix~\ref{app:embedding}. This choice guarantees the same mathematical structure for the propagators as for the free-particle (or the HF) propagator, and it is expected to provide an accurate description when the retarded/advanced embedding self-energy depends weakly on frequency around the Fermi level. In contrast, the frequency dependence of the lesser/greater embedding self-energy can be included in Eq.~\eqref{eq:collint-t}, cf.~Eq.~\eqref{eq:freq-integral}.

While Eq.~\eqref{eq:eom-rho} in combination with Eq.~\eqref{eq:gkba} [and the self-energies in Eqs.~\eqref{eq:sigma-hf}, \eqref{eq:sigma-2b}, and~\eqref{eq:sigma-em}] represents a closed set of equations, it only applies to the GKBA in the absence of initial contact and correlations, $\mathcal{I}^{\text{ic}}=0$ in Eq.~\eqref{eq:collint-ic}. This is because the GKBA does not provide an approximation for the mixed functions in Eq.~\eqref{eq:collint-ic}. It is possible to pass over this issue by starting the time-evolution from an initially noncorrelated state and then building up correlations by adiabatically switching on the many-particle and embedding effects~\cite{Rios2011, Hermanns2012}. This procedure may, however, lead to unpractically long propagation times putting the computational gain of the GKBA in jeopardy compared to the full Kadanoff--Baym equations~\cite{Karlsson2018, Tuovinen2019pssb, Tuovinen2020}. However, the inclusion of the initial correlations has been shown to be possible also within GKBA~\cite{Semkat2003, Karlsson2018, Hopjan2019, Bonitz2019}.

In Ref.~\onlinecite{Karlsson2018} a closed-form expression for $\mathcal{I}^{\text{ic}}$ in terms of $\rho$ was derived for a closed system where electron--electron interactions were described by the second-order Born self-energy. In the next section we will outline a similar procedure to evaluate $\mathcal{I}^{\text{ic}}(t)$ for the embedding self-energy. This procedure can directly be combined with many-body self-energies, making it possible to perform GKBA time-evolution for an initially correlated and contacted transport setup. This constitutes a partition-free framework for electronic transport in terms of the GKBA.

\section{Initial contacting collision integral}\label{sec:ic}

Let us start by separating the collision integral in Eq.~\eqref{eq:collint-ic} for the correlation and contacting contributions as
\be\label{eq:separate}
\mathcal{I}^{\text{ic}}(t) = \mathcal{I}_{\text{mb}}^{\text{ic}}(t) + \mathcal{I}_{\text{em}}^{\text{ic}}(t) .
\ee
For the first term (mb), we directly use the result derived in Ref.~\onlinecite{Karlsson2018} employing the second-order Born self-energy. For the second term (em), we take the self-energies as the embedding ones
\be\label{eq:icintegral}
\mathcal{I}_{\text{em}}^{\text{ic}}(t) \equiv -\im \int_0^\b \ud \tau \varSigma_{\text{em}}^\rceil(t,\tau) G^\lceil (\tau,t) .
\ee
The objects in Eq.~\eqref{eq:icintegral} satisfy the same analytic structure as in Ref.~\onlinecite{Karlsson2018}, enabling us to write a generalized fluctuation-dissipation theorem for the Green's function and self-energy. This further allows for writing Eq.~\eqref{eq:icintegral} equivalently in terms of the real-time lesser and greater functions~\cite{Karlsson2018}
\be\label{eq:collint}
\mathcal{I}_{\text{em}}^{\text{ic}}(t) = \int_{-\infty}^0 \ud \bar{t} [\varSigma_{\text{em}}^>(t,\bar{t}) G^<(\bar{t},t) - \varSigma_{\text{em}}^<(t,\bar{t})G^>(\bar{t},t)] ,
\ee
where Eqs.~\eqref{eq:gkba} and~\eqref{eq:freq-integral} are to be employed for the Green's functions and embedding self-energies, respectively. In general Eq.~\eqref{eq:collint} involves a convergence factor $\ex^{\eta \bar{t}}$ in the integrand (see Ref.~\onlinecite{Karlsson2018}). However, with contacted infinite leads this factor can be left out as the embedding self-energy accounts for proper convergence due to presence of a continuum of lead states. We notice that for $\bar{t} \in (-\infty,0)$ we have $\bar{t}<0<t$, i.e., the retareded Green function, $G^{\text{R}}(\bar{t},t) \propto \theta(\bar{t}-t)$, vanishes whereas the advanced Green function, $G^{\text{R}}(\bar{t},t) \propto \theta(\bar{t}-t)$, does not [cf.~Eq.~\eqref{eq:propagator}]. Importantly, for times $\bar{t}<0<t$, the single-particle density matrix is static, given by some equilibrium value, $\rho(\bar{t}) \equiv \rho^{\text{eq}}$. In this time interval, the HF Hamiltonian also becomes static, $h_{\text{HF}}(\bar{t}) = h_{\text{HF}}[\rho(\bar{t})] = h_{\text{HF}}[\rho^{\text{eq}}] \equiv h_{\text{HF}}^{\text{eq}}$. These time intervals may then be separated using the group property
\be\label{eq:ga}
G^{\text{A}}(\bar{t},t) = -\im G^{\text{A}}(\bar{t},0)G^{\text{A}}(0,t) = \ex^{-\im (h_{\text{HF}}^{\text{eq}}+\im \varGamma/2)\bar{t}} G^{\text{A}}(0,t) .
\ee
Note that here the equilibrium system is taken as coupled, and $\varGamma$ appearing in the exponent is due to the WBA, in accordance with Eq.~\eqref{eq:propagator}.

With these considerations Eq.~\eqref{eq:collint} can be expanded as
\begin{widetext}
\begin{align}
\mathcal{I}_{\text{em}}^{\text{ic}}(t) = \int_{-\infty}^0 \ud \bar{t} \Big\{ & -\im \sum_\a \ex^{-\im \psi_\a(t,\bar{t})} \int \frac{\ud \w}{2\pi} [1-f(\w-\mu)] \varGamma_\a(\w) \ex^{-\im\w(t-\bar{t})}\rho^{\text{eq}}  \ex^{-\im(h_{\text{HF}}^{\text{eq}}+\im \varGamma/2)\bar{t}} G^{\text{A}}(0,t) \nonumber \\
& +\im \sum_\a \ex^{-\im \psi_\a(t,\bar{t})} \int \frac{\ud \w}{2\pi} f(\w-\mu) \varGamma_\a(\w) \ex^{-\im\w(t-\bar{t})}(1-\rho^{\text{eq}}) \ex^{-\im(h_{\text{HF}}^{\text{eq}}+\im \varGamma/2)\bar{t}} G^{\text{A}}(0,t) \Big\} .
\end{align}
Here, it is worth noting that the frequency-dependence of $\varGamma_\a(\w)$ results from the lesser/greater embedding self-energy in Eq.~\eqref{eq:freq-integral}, which itself is of general form and does not require the WBA. For $\bar{t} \in (-\infty,0)$ the system is in equilibrium, i.e., external fields are not switched on. For the bias voltage phase factor we therefore have $\psi_\a(t,\bar{t}) = \psi_\a(t,0)$. Then, by canceling and combining some terms we may isolate the $\bar{t}$ integration
\be\label{eq:deriv}
\mathcal{I}_{\text{em}}^{\text{ic}}(t) = \im \sum_\a \ex^{-\im\psi(t,0)}\int\frac{\ud \w}{2\pi} \varGamma_\a(\w) [f(\w-\mu)-\rho^{\text{eq}}] \ex^{-\im \w t} \left[ \int_{-\infty}^0 \ud \bar{t} \ex^{\im (\w-h_{\text{HF}}^{\text{eq}}-\im\varGamma/2)\bar{t}} \right] G^{\text{A}}(0,t) .
\ee
The time integration is now straightforward to perform and we are left with a frequency integral only.
This gives us as the final result for the initial contacting collision integral
\be\label{eq:freqintegral}
\mathcal{I}_{\text{em}}^{\text{ic}}(t) = \sum_\a \ex^{-\im \psi_\a(t,0)}\int\frac{\ud \w}{2\pi} \varGamma_\a(\w) [f(\w-\mu) - \rho^{\text{eq}}] \frac{\ex^{-\im \w t}}{\w-(h_{\text{HF}}^{\text{eq}}+\im\varGamma/2)} G^{\text{A}}(0,t) ,
\ee
\end{widetext}
where we used the notation $c-A \equiv c\bunitmat - A$ and $1/A \equiv A^{-1}$ for a scalar $c$ and a matrix $A$.

Importantly, we are left with no time integrations, so Eq.~\eqref{eq:freqintegral} can be evaluated at any time $t$ with minor computational cost, provided that $G^{\text{A}}(0,t)$ is already available during the time evolution. We discuss in Appendix~\ref{app:contour} the case of taking explicitly the WBA for Eq.~\eqref{eq:freqintegral}, which lightens the computational cost even more. If, in addition, we consider a specific but frequently used harmonic bias voltage profile, $V_\a(t) = V_\a^0 + A_\a \cos (\varOmega_\a t +\phi_a)$, also the phase factor $\ex^{-\im\psi_\a(t,0)}$ can be expanded in terms of Bessel functions~\cite{Jauho1994,Ridley2015,Ridley2017}.

The time-dependent current between the molecular region and the leads can be calculated by the Meir--Wingreen formula~\cite{Meir1992,Jauho1994}
\begin{align}\label{eq:mw}
I_\a(t) = 4 \text{Re} \text{Tr} \int_{t_0}^t \ud \bar{t} \big[ & (\varSigma_{\text{em}}^>)_\a(t,\bar{t})G^<(\bar{t},t) \nonumber \\ 
- & (\varSigma_{\text{em}}^<)_\a(t,\bar{t})G^>(\bar{t},t) \big] .
\end{align}
This needs to be adjusted to include the effect from the initial contacting collision integral in Eq.~\eqref{eq:freqintegral}. The adjustment can be obtained by writing Eq.~\eqref{eq:freqintegral} as $\mathcal{I}_{\text{em}}^{\text{ic}}(t) = \sum_\a (\mathcal{I}_{\text{em}}^{\text{ic}})_\a(t)$, and then identifying the time-dependent current as
\be
I_\a(t) \longrightarrow I_\a(t) + 4 \text{Re} \text{Tr} (\mathcal{I}_{\text{em}}^{\text{ic}})_\a(t) .
\ee
We note that a corresponding contribution arising from the many-body self-energy vanishes due to conservation laws and the self-consistent solution to the equations of motion~\cite{Talarico2019}.

\section{Results}\label{sec:results}

We now demonstrate the protocol derived in the previous section. In all of the numerical simulations presented we consider two separate cases: (1) the standard GKBA time evolution where the correlated and contacted initial state is prepared by an adiabatic switching procedure for $t \in [-T,0]$ and then switching on the bias voltage at $t=0$; and (2) the GKBA time evolution supplemented with the initial correlations and contacting collision integral, starting the simulation directly at $t=0$ with the bias voltage. We refer to the former case as `GKBA$|$AS' and to the latter as `GKBA$|$IC'. As we wish to analyze the validity of the initial contacting protocol, we consider the electronic interactions at the HF and 2B level. Note that in the HF case $\mathcal{I}_{\text{mb}}^{\text{ic}}=0$ in Eq.~\eqref{eq:separate}, and in the 2B case this contribution is evaluated using the approach of Ref.~\onlinecite{Karlsson2018}.

We consider two different molecular junctions where the `molecule' being coupled to macroscopic metallic leads is (i) cyclobutadiene and (ii) a graphene nanoflake. The modeling for the molecular regions is done at the Pariser--Parr--Pople~\cite{PariserParr1953,Pople1953} (PPP) level, where the kinetic and interaction matrix elements are obtained semi-empirically by fitting to more sophisticated calculations.

The macroscopic metallic leads are described as noninteracting semi-infinite tight-binding lattices. The role of the leads is to act as particle reservoirs and as biased electrodes accounting for a potential drop across the molecular region. The potential drop is modeled by a symmetric bias voltage $V_L=-V_R \equiv V$, see~Appendix~\ref{app:details}. We also consider the zero-temperature limit at which we derive in Appendix~\ref{app:embedding} a fast and accurate analytical representation of the embedding self-energy in terms of Bessel and Struve functions. The matrix structure of the embedding self-energy is specified by the coupling matrix elements between the molecular region and the leads: In all of the cases considered the left-most atoms of the molecular region are coupled to the left lead, and the right-most atoms of the molecular region are coupled to the right lead with equal coupling strength $t_{\a C}$ where $\a=L,R$. The energy scale in the lead is specified by the hopping strength between the lead sites $t_\a$.

\subsection{Cyclobutadiene}

We consider a cyclobutadiene molecule attached to donor--acceptor-like leads. This is a circular molecule of $4$ atomic sites, and it is modeled by PPP parameters obtained by fitting to an effective valence shell Hamiltonian method~\cite{Martin1996}. The single-particle matrix is taken as (in atomic units)
\be
h=-\begin{pmatrix} 0.903 & 0.119 & 0 & 0.098 \\
  	  	           0.119 & 0.903 & 0.098 & 0 \\
		           0 & 0.098 & 0.903 & 0.119 \\
		           0.098 & 0 & 0.119 & 0.903 \end{pmatrix},
\ee
and the two-body interaction is of the form $v_{mnpq}=v_{mn}\delta_{mq}\delta_{np}$ with (in atomic units)
\be
v = \begin{pmatrix} 0.433 & 0.201 & 0.165 & 0.202 \\
            		0.201 & 0.430 & 0.202 & 0.165 \\
            		0.165 & 0.202 & 0.433 & 0.201 \\
            		0.202 & 0.165 & 0.201 & 0.430 \end{pmatrix} .
\ee
Note the slightly asymmetric structure of the hopping and interaction matrix elements due to cyclobutadiene being a rectangle, not a square~\cite{Breslow2008}. The coupling between the molecule and the leads is of equal strength $t_{\a C} = -0.06$~a.u. from the molecular sites $1$ and $4$ to the left lead and from the molecular sites $2$ and $3$ to the right lead. The hopping energy in the leads is $t_\a=-0.24$~a.u., which gives for the tunneling rate $\varGamma_\a = 2t_{\a C}^2/|t_\a| = 0.03$~a.u. The chemical potential is set between the highest occupied molecular orbital (HOMO) and the lowest unoccupied molecular orbital (LUMO), $\mu=-0.119$~a.u., of the isolated molecule with two electrons. Such simplified modeling of the molecular junction enables us to address the new approach with mathematical transparency, also comparing the GKBA with the full Kadanoff--Baym equation (KBE) approach of Ref.~\onlinecite{Myohanen2009}. 

\begin{figure}[t]
\centering
\includegraphics[width=0.49\textwidth]{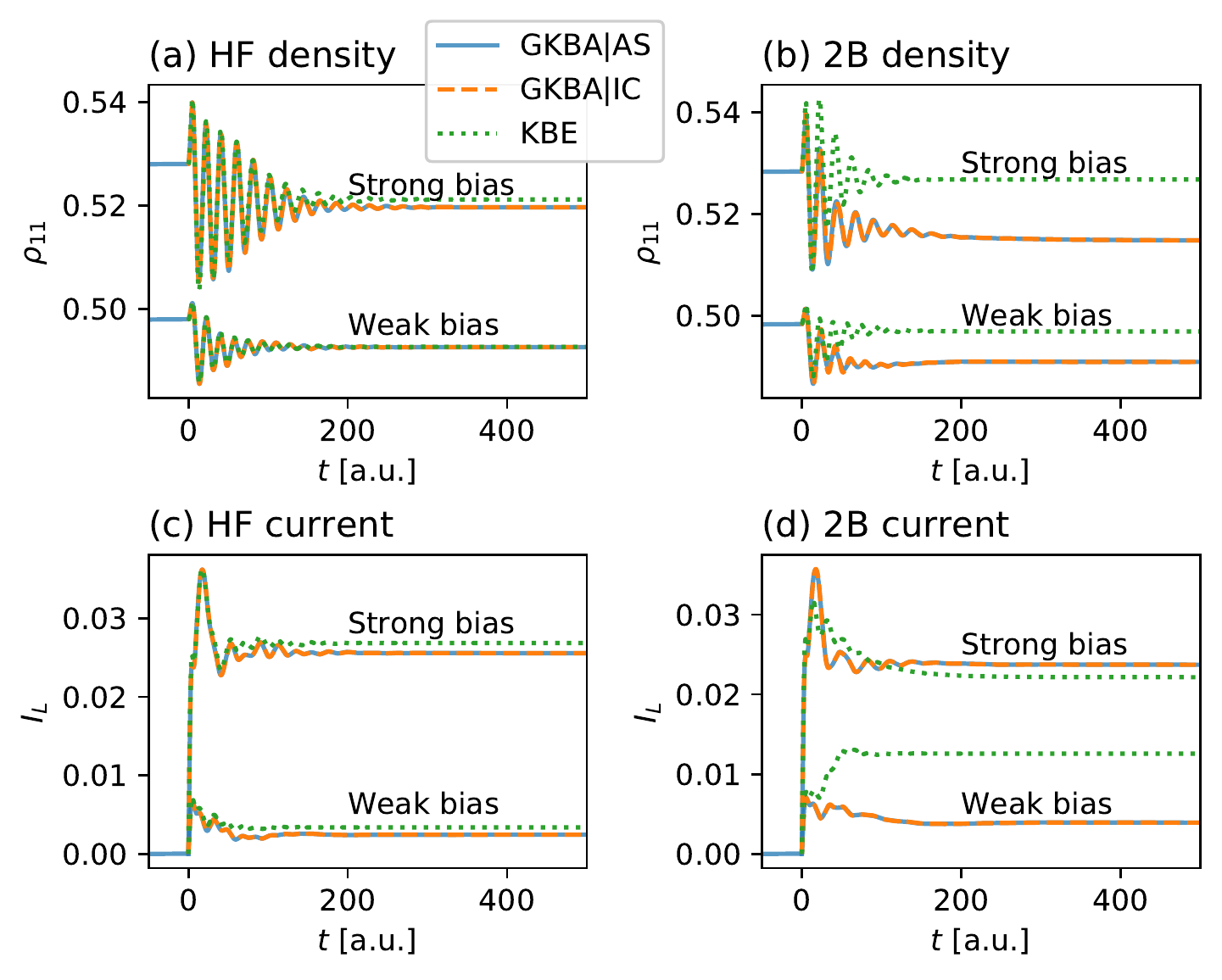}
\caption{Time-dependent densities (panels a and b) and currents (panels c and d) in a cyclobutadiene molecular junction under the influence of voltage biases $V=0.06$ (weak) and $V=0.24$ (strong) at the HF (panels a and c) and 2B (panels b and d) level. The GKBA time evolution with adiabatic switching is shown with the solid line and the switching time extends over the figure frame to $t=-T=-500$. The GKBA time evolution starting from $t=0$ with the initial contacting protocol is shown with the dashed line, and a benchmark solution to the full KBE is shown with the dotted line. The densities in panels (a) and (b) for the strong bias case are shifted upwards by $0.03$ for clarity.}
\label{fig:cyclobutadiene}
\end{figure}

\begin{figure}[t]
\centering
\includegraphics[width=0.49\textwidth]{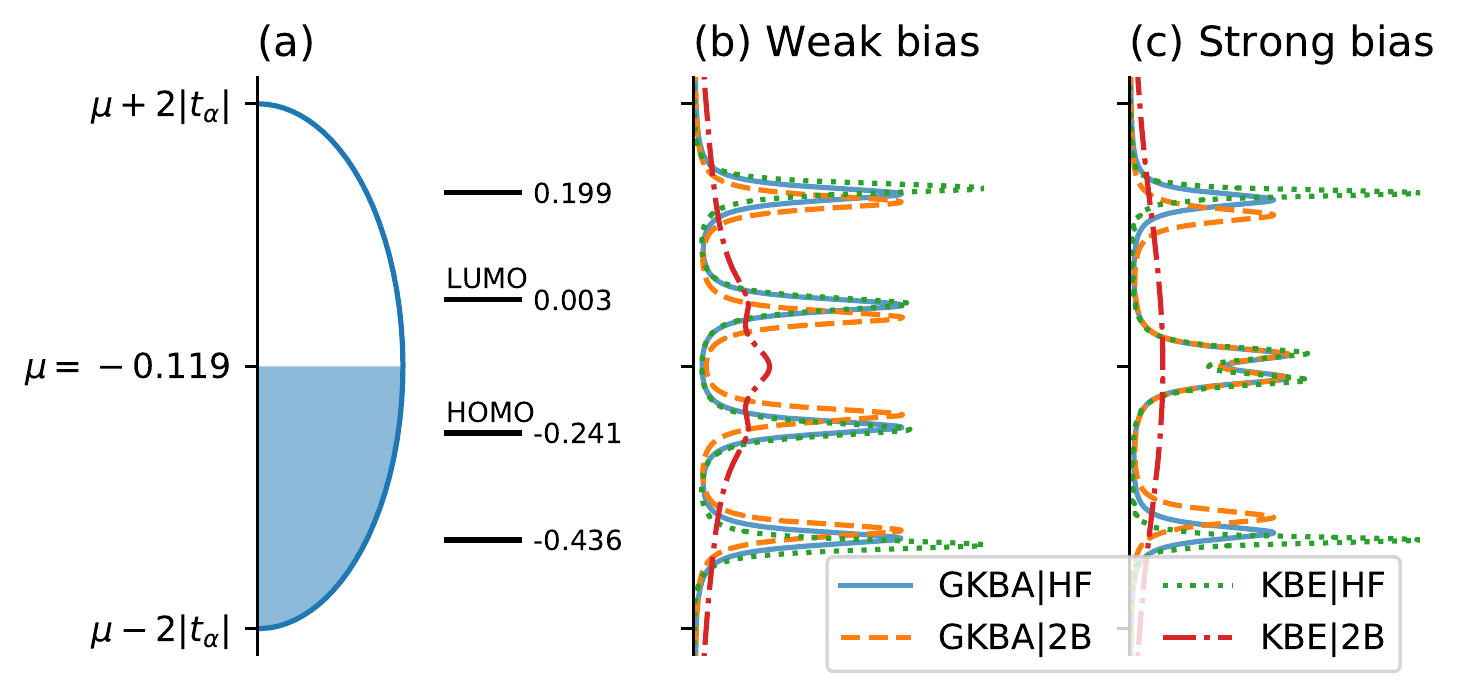}
\caption{Energy diagrams of the cyclobutadiene molecular junction. Panel (a) shows the energy-dependence (vertical axis) of the imaginary part of the embedding self-energy, i.e., the lead density of states. The HF energy levels of the isolated molecule are shown as horizontal lines. Panels (b) and (c) show the out-of-equilibrium spectral functions in the weak ($V=0.06$) and strong ($V=0.24$) bias case, respectively. The vertical energy axes of the spectral functions are aligned with the energy axis in panel (a).}
\label{fig:schematic}
\end{figure}

In Fig.~\ref{fig:cyclobutadiene} we show the time-dependent densities (at the first site) and currents (through the left lead interface) in the cyclobutadiene molecular junction. We see that the restart protocol of GKBA with the initial contacting (GKBA$|$IC) is very well in agreement with the adiabatic switching (GKBA$|$AS) for both the density and the current, and for both weak and strong bias. In addition, we have checked (not shown) that in the absence of bias the GKBA$|$IC time evolution remains stable and unchanged from the state described by $\rho^{\text{eq}}$.
Even though the bias window in the strong-bias case extends up to half the bandwidth, we still observe a satisfactory agreement between GKBA and KBE at the HF level. At the 2B level compared to full KBE, we find a typical mismatch of the steady-state density and current. This can be addressed in terms of the out-of-equilibrium spectral function, which is calculated as a Fourier transformation with respect to the relative-time coordinate $t_r \equiv t-t'$:
\begin{align}
A(\w) = \im \int \ud t_r \ex^{\im\w t_r} \mathrm{Tr}[ & G^>(T_c+t_r/2,T_c-t_r/2) \nonumber \\
- & G^<(T_c+t_r/2,T_c-t_r/2)] ,
\end{align}
where we set the center-of-time coordinate, $T_c\equiv (t+t')/2$, to half the total propagation time, so that the relative-time coordinate spans the maximal range diagonally in the two-time plane. In Fig.~\ref{fig:schematic} we show the energy diagram together with the out-of-equilibrium spectral functions. As the GKBA in Eq.~\eqref{eq:gkba} satisfies the exact condition $G^> - G^< = G^{\mathrm{R}} - G^{\mathrm{A}}$, the GKBA spectral function adheres to the form of the HF propagators in Eq.~\eqref{eq:propagator}. This is generally in agreement with Refs.~\onlinecite{Thygesen2008, Myohanen2009} for similar systems. Here the interaction is relatively strong, so the KBE 2B spectral function is completely smeared out.

\subsection{Graphene nanoflake}

\begin{figure}[t]
\centering
\includegraphics[width=0.45\textwidth]{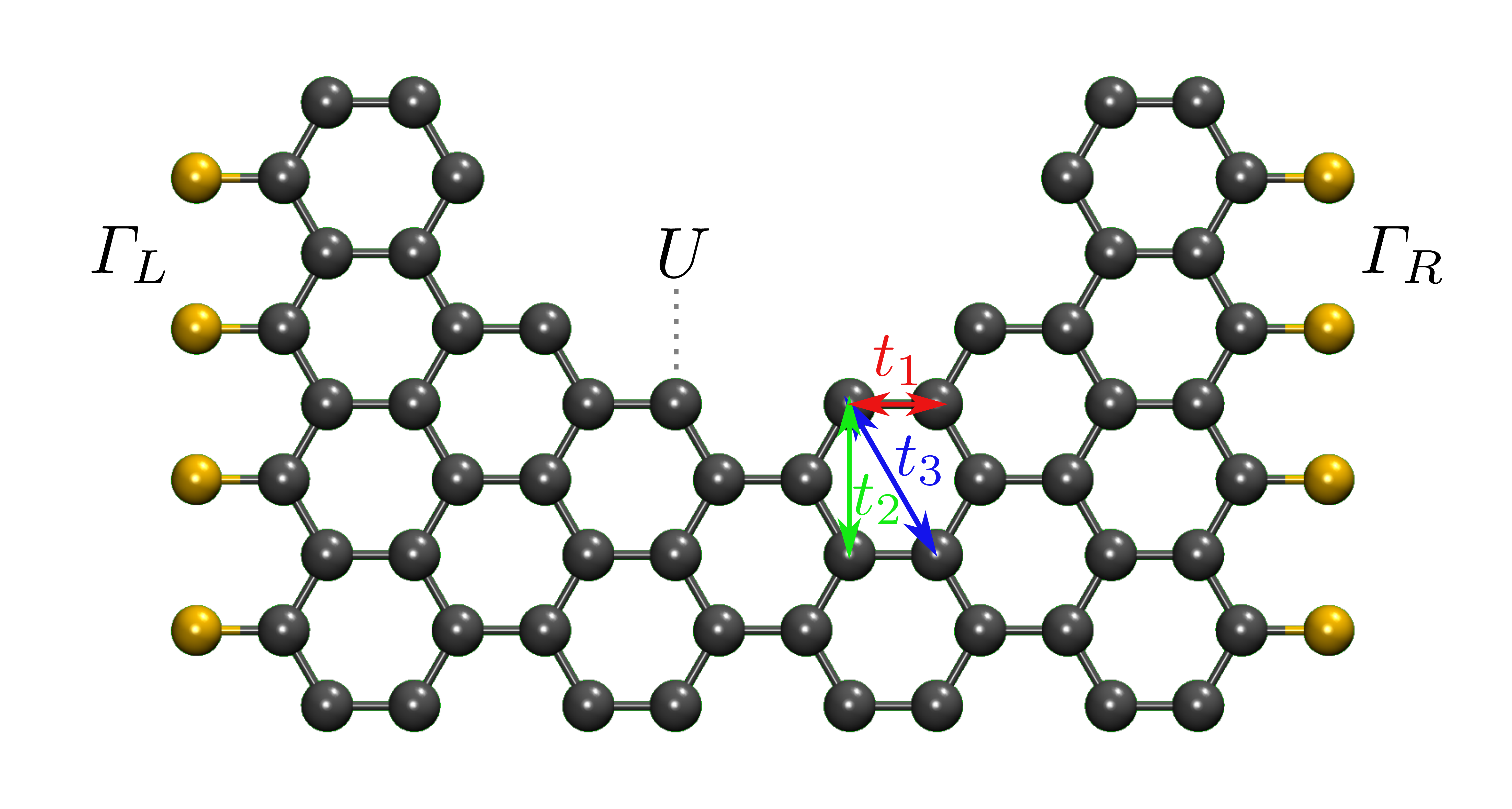}
\caption{Graphene nanoflake molecular junction. The left-most carbon atoms (black) are connected to the left lead (yellow) with tunneling rate $\varGamma_L$, and the right-most carbon atoms (black) are connected to the right lead (yellow) with tunneling rate $\varGamma_R$. Only the terminal sites of the leads are depicted. Red, green, and blue arrows signify the hopping energies between first-, second-, and third-nearest neighbors, respectively. Electron--electron interaction is of Hubbard type with strength $U$.}
\label{fig:notched}
\end{figure}

We then consider a graphene nanoflake which has more structural complexity, see Fig.~\ref{fig:notched}. The molecular region is a notched armchair graphene nanoribbon of width $9$ and length $8$ similar to Ref.~\onlinecite{Hancock2010} where a generalized tight-binding model was proposed. It was found that a single parameter set with first-, second-, and third-nearest neighbor hoppings $t_1=-2.7$~eV, $t_2=-0.20$~eV, $t_3=-0.18$~eV, and a Hubbard interaction $U=2.0$~eV accurately reproduced density-functional theory based results for both the band structure and the conductance. We choose these values as the PPP parameters along the hexagonal lattice in Fig.~\ref{fig:notched} for Eq.~\eqref{eq:hamiltonian}: $h_{mn}=t_1,t_2,t_3$ for first-, second-, and third-nearest neighbors, respectively, and $v_{mnpq}=U\delta_{mn}\delta_{mp}\delta_{mq}$. Here we describe the interactions at the 2B level. We note that, as we include next-nearest neighbor hoppings, the electron-hole symmetry is not preserved~\cite{CastroNeto2009}. In addition, the nanoflake is coupled to the left and right leads from the left-most and right-most carbon atoms, respectively (see~Fig.~\ref{fig:notched}). We consider two cases of tunneling rates between the graphene nanoflake and the leads $\varGamma_\a \in \{0.02|t_1|,0.2|t_1|\}$, and we fix the bias voltage to $V=|t_1|$ with respect to the chemical potential $\mu=1.44$~eV, which is set in the middle of the HOMO--LUMO gap of the isolated graphene nanoflake. As the energies are in electronvolts, $\varepsilon=1$~eV, we convert the units for time to seconds by $t=\hbar/\varepsilon\approx 6.582\cdot 10^{-16}$~s, and the units for current to amperes by $I = e\varepsilon/\hbar\approx 2.434\cdot 10^{-4}$~A.

\begin{figure}[t]
\centering
\includegraphics[width=0.475\textwidth]{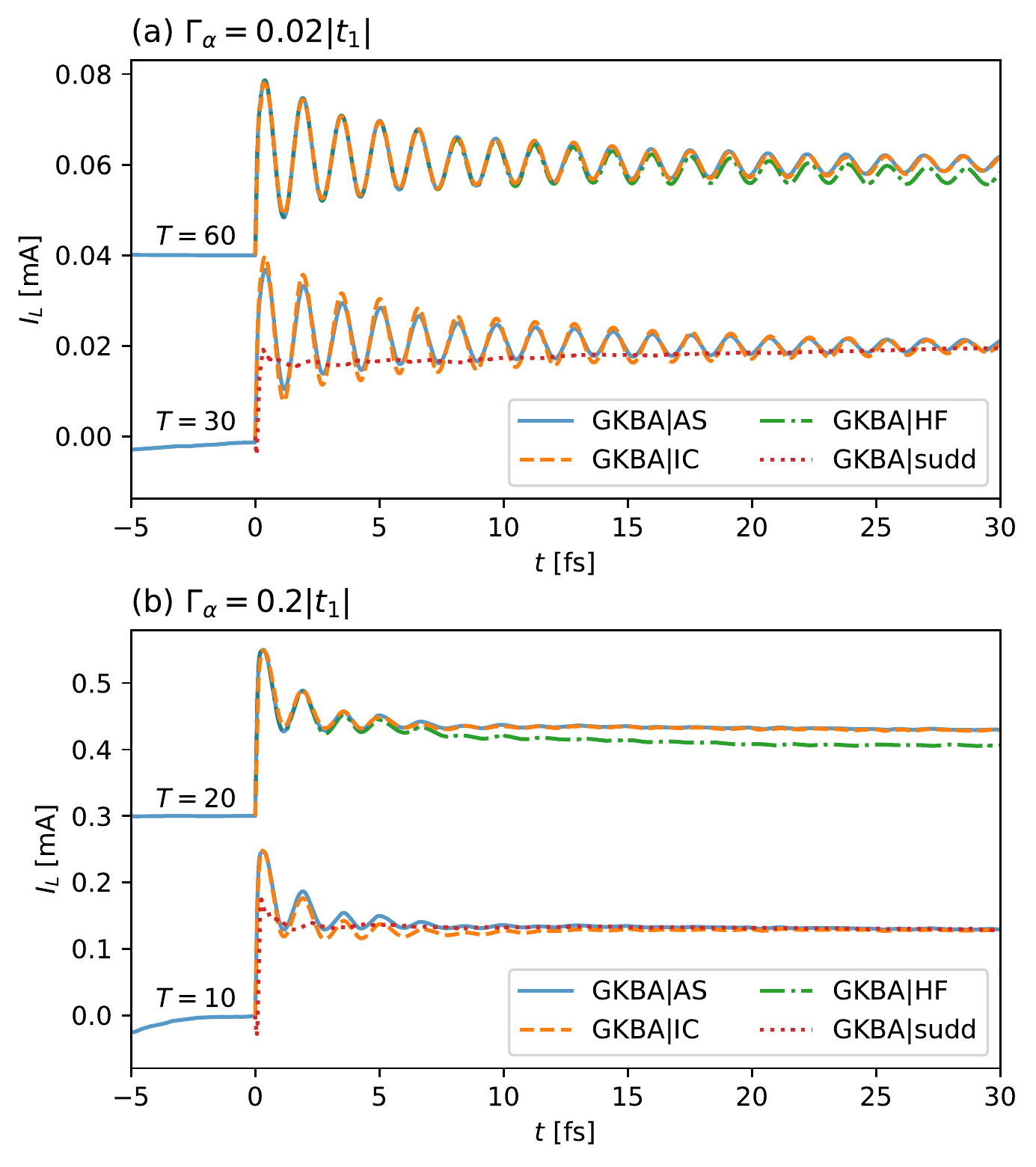}
\caption{Time-dependent currents through the left lead interface of a graphene nanoflake in (a) weak and (b) intermediate coupling regime. The GKBA time evolution with the 2B self-energy and with adiabatic switching (GKBA$|$AS) is shown with the solid line and the indicated switching time $t=-T$ extends over the figure frame. The GKBA time evolution with the 2B self-energy starting from $t=0$ with initial contact and correlations (GKBA$|$IC) is shown with the dashed line. A comparative time evolution with the HF self-energy is shown with the dash-dotted line. A sudden switch-on of both correlations (2B) and voltage at $t=0$ (GKBA$|$sudd) is shown with a dotted line. For clarity, we apply an upward shift of $0.04$~mA for the $T=60$~fs case in panel (a) and $0.3$~mA for the $T=20$~fs case in panel (b).}
\label{fig:graphene}
\end{figure}

The restart protocol with the initial contacting (GKBA$|$IC) relies on a converged initial state. In Fig.~\ref{fig:graphene} we show the time-dependent currents (through the left lead interface) for the graphene nanoflake molecular junction, and we study the role of the switching time $t=-T$ and the tunneling rate $\varGamma_\a$. First, the initial contacting (GKBA$|$IC) is in excellent agreement with the adiabatic switching (GKBA$|$AS) for both weak and intermediate coupling. Second, the relaxation time of the switching procedure is longer for weaker coupling~\cite{Latini2014}. This is reflected on the initial contacting protocol, which requires an equilibrium density matrix $\rho^{\text{eq}}$ as input. If this $\rho^{\text{eq}}$ does not result from a properly converged calculation, then GKBA$|$IC starts deviating from GKBA$|$AS, as can be seen in Fig.~\ref{fig:graphene} cases $T=10$~fs and $T=30$~fs.
We emphasize that the computational cost for the GKBA$|$IC is independent on how long the preparation stage takes: With one prepared $\rho^{\text{eq}}$, as many out-of-equilibrium simulations as desired may be performed (e.g. voltage sweep).

In Fig.~\ref{fig:graphene} we also show a simulation with the HF self-energy: Compared to the 2B case, the transient oscillation is roughly similar and only the steady-state value is affected within the GKBA description. For this size of the graphene structure and this model of interaction, this is reasonable as monolayer graphene devices are known to have fairly large coherent transport lengths~\cite{CastroNeto2009}. In Fig.~\ref{fig:graphene} we also show, for comparison, an ill-advised simulation of a simultaneous and sudden switch-on of both many-body correlations and contacts with voltage at $t=0$. Clearly, the transient features are completely misrepresented in this case, but the long-time limit coincides with the AS and IC results due to loss of memory of the initial state~\cite{Stefanucci2004}. Generally, with the chosen parameters for voltage and coupling, we find the absolute values of the stationary currents in the $10 \ \mu\mathrm{A}\div 1$~mA range, and the transient signature characterized in the $1 \div 100$~fs temporal range.

\begin{figure}[t]
\centering
\includegraphics[width=0.44\textwidth]{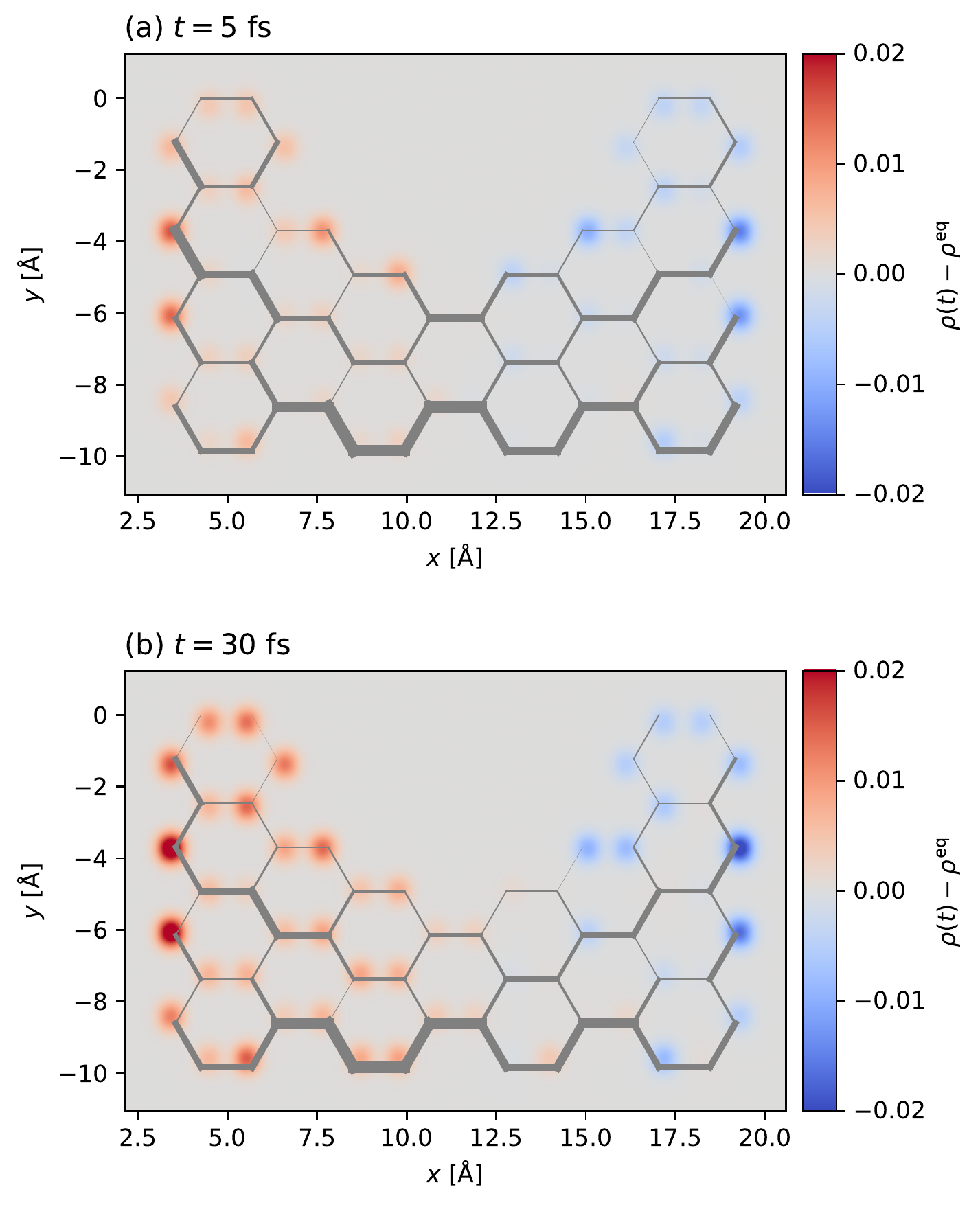}
\caption{Temporal snapshots of spatial charge-density variation (colormap) along the graphene nanoflake with respect to the equilibrium density. Relative strength of the nearest-neighbor bond current is indicated by the line thickness along the hexagonal lattice. The calculation corresponds to Fig.~\ref{fig:graphene}(a) with weak coupling $\varGamma_\a=0.02|t_1|$. Panel (a) shows the initial transient at $t=5$~fs while panel (b) is closer to the stationary state at $t=30$~fs.}
\label{fig:density}
\end{figure}

The dominant transient oscillation observed in Fig.~\ref{fig:graphene} is independent of the tunneling rate $\varGamma_\a$ and corresponds to a frequency of $|t_1|$ which is equal to the applied bias voltage. Therefore, these oscillations represent transitions between the biased Fermi level of the leads and the zero-energy states of the graphene nanoflake~\cite{Tuovinen2013, Tuovinen2014}. These zero-energy states correspond to surface states along the zigzag segments of the graphene nanoflake~\cite{Fujita1996,Tao2011}. This is confirmed in Fig.~\ref{fig:density} where the excited zero-energy modes are spatially focused along the surface during the initial transient. Interestingly, this effect therefore seems to be robust against electronic interactions. While it is outside of the scope of the present work, this effect could be associated with nontrivial spin polarization or antiferromagnetic alignment at the edges of the system~\cite{Yamashiro2003, Fernandez2008, Lado2015}. As the system relaxes towards the stationary state, the density response becomes more delocalized along the nanoflake. In Fig.~\ref{fig:density} we also show the relative strength of the bond current between nearest-neighbor carbon atoms. Due to the notched geometry, the current is appreciably stronger in the lower part of the nanoflake. We suspect larger graphene nanostructures would show even more pronounced separation of the charge and current density between the bulk and the surface during the initial transient~\cite{Rocha2015}.

\section{Conclusion}\label{sec:concl}

We have extended the GKBA approach for open quantum systems to a partition-free setting in electronic transport. We formulated the initial state for a molecular junction, before applying, e.g., a bias voltage driving, as correlated \emph{and} contacted. In practice, the contacted initial state was resolved as a separate calculation, which could be included to the out-of-equilibrium calculation via the initial contacting collision integral with a minor computational cost. This approach could directly be combined with the initial correlations in Ref.~\onlinecite{Karlsson2018}, making it possible to perform GKBA time-evolution for an initially correlated and contacted transport setup.

The inclusion of the initial contacting collision integral is very general. Since it only concerns the embedding self-energy, extensions to more sophisticated correlation self-energies, such as the $T$-matrix~\cite{Galitskii1958} or the $GW$ approximation~\cite{Hedin1965}, are directly applicable. In addition, extensions to correlated approximations to the propagators are applicable as long as they can be represented by $G^{\text{R}/\text{A}}(t,t') \propto \exp\{-\im \int_{t'}^t \ud \bar{t} [h_{\text{qp}}(\bar{t})\mp\im\varGamma/2]\}$, where $h_{\text{qp}}\equiv h_{\text{HF}}+\widetilde{\varSigma}$ includes some quasiparticle effects~\cite{Latini2014,Perfetto2018}. Even though we only considered constant bias voltage profiles, the driving could also be modulated in time~\cite{Jauho1994,Ridley2015,Ridley2017}.

We demonstrated the developed method by studying transient current signatures in carbon-based molecular junctions. A compact system of a cyclobutadiene molecule allowed for a transparent comparison of the new approach to adiabatic switching GKBA, and even to the full Kadanoff--Baym equation. Time-resolved densities and currents via the initial contacting approach were found to be in excellent agreement with the adiabatic switching approach in both weak- and strong-bias regimes. A more structurally complex system of a graphene nanoflake addressed both the potential and the limitations of the new approach. While the comparison with the adiabatic switching approach was found to be successful, it was important to verify the convergence of the initially correlated and contacted state before initiating an out-of-equilibrium simulation. We also found predominant oscillations in the transient current signal associated with virtual transitions between the graphene surface states and the biased Fermi levels of the lead. Together with recent experimental developments in ultrafast techniques these findings highlight the potential of addressing transiently emerging topological phenomena in molecular junctions out of equilibrium.

\begin{acknowledgments}
This research was funded by the Academy of Finland projects 321540 (R.T.) and 317139 (R.v.L.).
G.S. and E.P. acknowledge financial support from MIUR PRIN Grant No. 20173B72NB, from INFN through the TIME2QUEST project and from Tor Vergata University through the Beyond Borders Project ULEXIEX.
The computer resources of the Finnish IT Center for Science (CSC) and the FGCI project (Finland) are acknowledged.
R.T. also acknowledges useful discussions with N. W. Talarico, F. Cosco, and N. Lo Gullo.
\end{acknowledgments}

\section*{Data availability}
The data that support the findings of this study are available from the corresponding author upon reasonable request.

\appendix

\section{Background for the NEGF-GKBA equations}\label{app:details}

Generally, the Hamiltonian in Eq.~\eqref{eq:hamiltonian} is described with an argument $z$ referring to a time parameter on the Konstantinov--Perel' time contour~\cite{Konstantinov1961} $\g \equiv \g_- \oplus \g_+ \oplus \g_{\text{M}} \equiv (t_0,t) \oplus (t,t_0) \oplus (t_0,t_0 - \im \beta)$, where $t_0$ marks the beginning of a transport process generated by, e.g., a voltage switch-on, $t$ is the observation time, and $\beta=1/(k_{\text{B}}T)$ is the inverse temperature. The molecular junction Hamiltonian may then be specified for all contour times as~\cite{Ridley2018}
\begin{align}
\eps_{k\a}(z) & = \begin{cases}\eps_{k\a} + V_\a(t) & \text{when} \ z \in \g_- \oplus \g_+ \\ \eps_{k\a}-\mu & \text{when} \ z \in \g_{\text{M}} , \end{cases} \label{eq:bias}\\
h_{mn}(z) & = \begin{cases}h_{mn} + u_{mn}(t) & \text{when} \ z \in \g_- \oplus \g_+ \\ h_{mn}-\mu\delta_{mn} & \text{when} \ z \in \g_{\text{M}} , \end{cases} \label{eq:gate} 
\end{align}
where we introduced a bias voltage profile $V_\a(t)$ for the lead energy dispersion, a nonlocal potential profile $u_{mn}(t)$ for the molecular device, and the equilibrium chemical potential $\mu$. The coupling $T_{mk\a}(z)$ and interaction $v_{mnpq}(z)$ matrix elements can either be set equal and nonzero for all $z$, or zero for $z\in\g_{\text{M}}$ and proportional to a switching function $f(t)$ for $z\in\g_-\oplus\g_+$ on the horizontal branches.

The one-electron Green's function is defined on the time contour as~\cite{svlbook}
\be
G_{xy}(z,z') = -\im \langle \mathcal{T}_\g [ \hat{c}_{x}(z) \hat{c}_y^\dagger(z')] \rangle ,
\ee
where $\mathcal{T}_\g$ is the contour-time-ordering, the creation and annihilation operators are represented in the Heisenberg picture, and the ensemble average $\langle \cdot \rangle$ is taken as a trace over the density matrix. The Green's function satisfies the equation of motion (in matrix form)~\cite{svlbook}
\be
[\im \partial_z \bunitmat - \bH(z)]\bG(z,z') = \delta(z,z')\bunitmat + \int_\g \ud \bar{z} \bS_{\text{mb}}(z,\bar{z})\bG(\bar{z},z) , \label{eq:eom1}
\ee
and the corresponding adjoint equation. In Eq.~\eqref{eq:eom1} we introduced a block matrix structure with respect to the basis of single-electron states
\be
\bH(z) = \begin{pmatrix} h_{\a\a'}(z) & h_{\a C}(z) \\ h_{C\a'}(z) & h_{CC}(z) \end{pmatrix} ,
\ee
where the lead part is diagonal, $(h_{\a\a'})_{kk'}(z) = \delta_{\a\a'}\delta_{kk'}\eps_{k\a}(z)$, the tunneling is through the molecular device, $(h_{C\a})_{mk}(z) = T_{mk\a}(z)$, and $(h_{CC})_{mn}(z)=h_{mn}(z)$. In Eq.~\eqref{eq:eom1} we also wrote the many-body self-energy $\bS_{\text{mb}}$ accounting for the electronic interactions. While this interaction is constricted to the molecular region only, the Green's function matrix has nonzero entries everywhere:
\be
\bS_{\text{mb}} = \begin{pmatrix} 0 & 0 \\ 0 & (\varSigma_{\text{mb}})_{CC} \end{pmatrix} , \qquad \bG = \begin{pmatrix} G_{\a\a'} & G_{\a C} \\ G_{C\a'} & G_{CC} \end{pmatrix} .
\ee
The integration in Eq.~\eqref{eq:eom1}
is performed over the Konstantinov--Perel' contour through the Langreth rules~\cite{Langreth1972, Langreth1976}. In this procedure, the contour-time functions are represented in real-time components: lesser ($<$), greater ($>$), retarded (R), advanced (A), left ($\lceil$), right ($\rceil$), and Matsubara (M) depending on the contour-time arguments~\cite{svlbook}. 

We now consider the molecular region $C$ and take the projection of the equation of motion~\eqref{eq:eom1} onto these states. This procedure leads to~\cite{Myohanen2009}
\begin{align}\label{eq:eom-cc}
& [\im\partial_z - h_{CC}(z)]G_{CC}(z,z') = \delta(z,z') \nonumber \\
& + \int_\g \ud \bar{z} [(\varSigma_{\text{mb}})_{CC}(z,\bar{z})+(\varSigma_{\text{em}})_{CC}(z,\bar{z})]G_{CC}(\bar{z},z') ,
\end{align}
and a similar adjoint equation. In Eq.~\eqref{eq:eom-cc} we defined the embedding self-energy as
\be\label{eq:sigma-em}
(\varSigma_{\text{em}})_{CC}(z,z') = \sum_\a h_{C\a}(z)g_{\a\a}(z,z')h_{\a C}(z'),
\ee
where the Green's function of the noninteracting lead $g_{\a\a}$ satisfies $[\im\partial_z -h_{\a\a}(z)]g_{\a\a}(z,z') = \delta(z,z')$. As we are mainly considering the dynamical quantities within the molecular region, we will drop the $CC$ subscript for simplicity.

We consider the electronic interaction at the Hartree--Fock (HF) and second-order Born (2B) level, which are time-local and time-nonlocal, respectively
\be
\varSigma_{\text{mb}}(z,z') = \varSigma_{\text{HF}}(z)\delta(z,z') + \varSigma_{\text{2B}}(z,z') .
\ee
This separation allows us to remove the time-local part from the collision integral on the right-hand side of Eq.~\eqref{eq:eom-cc}, and couple it with the single-particle Hamiltonian on the left-hand side. Using the basis of single-electron states, the HF self-energy reads
\be\label{eq:sigma-hf}
(\varSigma_{\text{HF}})_{ij}(z) = \sum_{mn}[2v_{imnj}(z)\rho_{nm}(z) - v_{imjn}(z)\rho_{nm}(z)],
\ee
where $\rho(z) \equiv -\im G(z,z^+)$ is the single-particle density matrix. The 2B self-energy takes the form
\begin{align}\label{eq:sigma-2b}
& (\varSigma_{\text{2B}})_{ij}(z,z') \nonumber \\
& = \sum_{\mathclap{mnpqrs}} v_{irpn}(z)v_{mqsj}(z') \left[ 2G_{nm}(z,z')G_{pq}(z,z')G_{sr}(z',z) \right. \nonumber \\
& \left. \hspace{60pt} - G_{nq}(z,z')G_{pm}(z,z')G_{sr}(z',z)\right] ,
\end{align}
where the summation over the basis states can be reorganized for efficient computation~\cite{Perfetto2019, Schluenzen2019, Tuovinen2019}. 

Taking the equal-time limit, $z=t_-$, $z'=t_+$, in Eq.~\eqref{eq:eom-cc} and its adjoint, and employing the Langreth rules we obtain the GKBA equation of motion~\eqref{eq:eom-rho} for the single-particle density matrix in main text.

\section{Embedding self-energy}\label{app:embedding}
Since the leads are treated as noninteracting, they can be incorporated non-perturbatively into the equation of motion~\eqref{eq:eom-cc} using the embedding self-energy in Eq.~\eqref{eq:sigma-em}. On the real-time branch the relevant lead Green's functions are~\cite{svlbook}
\begin{align}
g_{k\a}^{\text{R}/\text{A}}(t,t') & = \mp \theta[\pm (t-t')]\ex^{-\im\int_{t'}^t \ud \bar{t} [\eps_{k\a}+V_\a(\bar{t})]} , \\
g_{k\a}^\lessgtr(t,t') & = \pm \im f[\pm(\eps_{k\a}-\mu)]\ex^{-\im\int_{t'}^t\ud \bar{t}[\eps_{k\a}+V_\a(\bar{t})]} ,
\end{align}
where $f(x)=1/(1+\ex^{\beta x})$ is the Fermi function at inverse temperature $\beta$ with the property $f(-x)=1-f(x)$. The retarded and advanced embedding self-energies are then given by~\cite{Ridley2015}
\be
(\varSigma_{\text{em}}^{\text{R}/\text{A}})_\a(t,t') = \ex^{-\im\psi_\a(t,t')}\int\frac{\ud\w}{2\pi}[\varLambda_\a(\w) \mp \im\varGamma_\a(\w)/2] ,
\ee
where $\psi_\a(t,t')\equiv \int_{t'}^t \ud \bar{t} V_\a (\bar{t})$ is the phase factor originating from the bias voltage profile, and we wrote the level-shift and level-width matrices as
\begin{align}
(\varLambda_\a)_{mn}(\w) & = \sum_k T_{mk\a} \mathcal{P}\left(\frac{1}{\w-\eps_{k\a}}\right)T_{k\a n} , \label{eq:shift}\\
(\varGamma_\a)_{mn}(\w) & = 2\pi\sum_k T_{mk\a}\delta(\w-\eps_{k\a})T_{k\a n} , \label{eq:width}
\end{align}
respectively. In Eqs.~\eqref{eq:shift} and~\eqref{eq:width} we used $1/(\w-\eps_{k\a}\pm \im\eta) = \mathcal{P}(1/(\w-\eps_{k\a}))\mp\im\pi\delta(\w-\eps_{k\a})$ with $\eta$ being a positive infinitesimal and $\mathcal{P}$ denoting the principal value.

In the wide-band approximation (WBA) the level-width is taken as independent of frequency, $\varGamma_\a(\w) \approx \varGamma_\a$. This amounts to approximating the lead density of states being practically featureless in the energy scale of the molecular system. In this approximation the level-shift matrix vanishes due to Kramers--Kronig relations, and the retarded/advanced embedding self-energy becomes time-local
\begin{align}
\varSigma_{\text{em}}^{\text{R}/\text{A}}(t,t') & \equiv \sum_\a(\varSigma_{\text{em}}^{\text{R}/\text{A}})_\a(t,t') = \mp \im \sum_\a\varGamma_\a \delta(t-t') / 2 \nonumber \\
& = \mp \im \varGamma \delta(t-t') / 2 .
\end{align}
In a similar manner we obtain the lesser/greater embedding self-energy as~\cite{Croy2009}
\begin{align}\label{eq:freq-integral}
& (\varSigma_{\text{em}}^\lessgtr)_\a(t,t') \nonumber \\
& = \pm\im \ex^{-\im\psi_\a(t,t')}\int \frac{\ud \w}{2\pi}f[\pm(\w-\mu)]\varGamma_\a(\w)\ex^{-\im\w(t-t')} .
\end{align}
Even though we set ourselves in the regime where the WBA holds, in Eq.~\eqref{eq:freq-integral} we keep the frequency dependency of $\varGamma_\a$ to ensure convergence of the frequency integral.

The matrix structure of the embedding self-energy is determined by the corresponding structure of the level-width matrix $\varGamma_\a$ which is specified by the coupling and lead Hamiltonians in Eq.~\eqref{eq:width}. We now address the frequency integral in Eq.~\eqref{eq:freq-integral} and consider the effective form of the level-width matrix for a one-dimensional semi-infinite tight-binding lead~\cite{Myohanen2009}
\be
\varGamma_\a(\w) \propto \sqrt{1-\left(\frac{\w-a_\a}{2|t_a|}\right)^2},
\ee
where $a_\a$ is the on-site energy of the sites in lead $\a$ and $t_\a$ the hopping energy between the sites in lead $\a$. This form also limits the integration range to $\w \in [a_\a-2|t_\a|,a_\a+2|t_\a|]$, see~Fig.~\ref{fig:schematic}(a). We also consider the zero-temperature limit at which the Fermi function becomes a Heaviside step function, $f(\w-\mu) \to \theta(\mu-\w)$, and this introduces a further cutoff to the integral. We also write $t_r=t-t'$ for brevity and obtain
\begin{align}
& \int_{-\infty}^\infty \frac{\ud \w}{2\pi}f(\w-\mu)\varGamma_\a(\w)\ex^{-\im\w(t-t')} \nonumber \\
\propto & \int_{a_\a-2|t_\a|}^\mu \frac{\ud\w}{2\pi}\sqrt{1-\left(\frac{\w-a_\a}{2|t_\a|}\right)^2}\ex^{-\im\w t_r} .
\end{align}
We make the identification that the on-site energies in the leads are aligned with the equilibrium chemical potential, $\mu=a_\a$, resulting in half filling for the leads' energy continua. Making a change of variables $x\equiv (\w-a_\a)/(2|t_\a|)$ we obtain
\begin{align}\label{eq:sqrt-integral}
& \int_{a_\a-2|t_\a|}^\mu \frac{\ud\w}{2\pi}\sqrt{1-\left(\frac{\w-a_\a}{2|t_\a|}\right)^2}\ex^{-\im\w t_r} \nonumber \\
= & \frac{|t_\a|}{\pi}\ex^{-\im a_\a  t_r} \int_{-1}^0 \ud x \sqrt{1-x^2}\ex^{-2 \im x |t_\a| t_r} \nonumber \\
= & \frac{\ex^{-\im a_\a  t_r}}{4 t_r}[J_1(2|t_\a| t_r)+\im H_1(2|t_\a| t_r)] ,
\end{align}
where on the last line we used an integral representation of the Bessel and Struve functions of the first kind~\cite{NIST:DLMF,Zhu2005}. We remind that the final result for the lesser embedding self-energy at the zero-temperature limit, obtained by inserting Eq.~\eqref{eq:sqrt-integral} in Eq.~\eqref{eq:freq-integral}, needs to be supplemented with the appropriate matrix structure. We also note that at the equal-time limit, $ t_r \to 0$, Eq.~\eqref{eq:sqrt-integral} reduces to a value of $|t_\a|/4$. The case of the greater embedding self-energy with $f[-(\w-\mu)]$ in Eq.~\eqref{eq:freq-integral} results in the integral $\int_0^1 \ud x \sqrt{1-x^2}\ex^{-2 \im x |t_\a| t_r}$ which has otherwise the same representation as in Eq.~\eqref{eq:sqrt-integral} but the sign in front of the Struve function is changed. The Struve function can be evaluated using an approximate expansion in terms of the Bessel functions~\cite{Aarts2003,Aarts2016}, or as a direct combination of power series and continued fraction~\cite{Zhang1996}.

\section{Evaluation of the frequency integral at the wide-band approximation}\label{app:contour}

In practice, we evaluated Eq.~\eqref{eq:freqintegral} by numerical integration for all the simulations presented. However, we can make some further analytical progress by taking explictly the WBA. We show here how, in this case, the frequency integral in Eq.~\eqref{eq:freqintegral} can further be expressed in terms of the hypergeometric function~\cite{NIST:DLMF} which can be evaluated using a fast and accurate numerical algorithm~\cite{Michel2008}. Now, $\varGamma_\a(\w)$ due to the lesser/greater embedding self-energy is also taken as independent of frequency, and Eq.~\eqref{eq:freqintegral} can be written as
\be\label{eq:freqintegral-wba}
\mathcal{I}_{\text{em}}^{\text{ic}}(t) = \sum_\a \varGamma_\a \ex^{-\im\psi_\a(t,0)}\int\frac{\ud\w}{2\pi} \frac{f(\w-\mu)\ex^{-\im\w t}}{\w-(h_{\text{HF}}^{\text{eq}}+\im\varGamma/2)}G^{\text{A}}(0,t).
\ee
Here we used the fact that the frequency integral is performed over the full real axis $\w \in (-\infty,\infty)$, and it can be evaluated using contour integration techniques. The exponential factor in the numerator converges only in the lower-half of the complex plane. In this region the contribution $\sim \rho^{\text{eq}}\ex^{-\im \w t}/(\w-h_{\text{HF}}^{\text{eq}}-\im\varGamma/2)$ from Eq.~\eqref{eq:freqintegral} does not contain any poles, so this contribution to the integral vanishes. The other contribution contains the Matsubara poles of the Fermi function, in the lower-half of the complex plane, keeping the integral nonzero.

We may then expand the result in the (right) eigenvector basis of the nonhermitian matrix
\be
(h_{\text{HF}}^{\text{eq}}+\im\varGamma/2) | \psi_j^{\text{R}} \rangle = \eps_j | \psi_j^{\text{R}} \rangle .
\ee
We recall that the left/right eigenvectors of a nonhermitian matrix form a biorthogonal basis set. The frequency integral of Eq.~\eqref{eq:freqintegral-wba} in this basis reads
\begin{align}
& \langle \psi_j^{\text{R}} | \int\frac{\ud \w}{2\pi} f(\w-\mu)\frac{\ex^{-\im \w t}}{\w-(h_{\text{HF}}^{\text{eq}}+\im\varGamma/2)} | \psi_k^{\text{R}} \rangle \nonumber \\
& = \langle \psi_j^{\text{R}} | \psi_k^{\text{R}} \rangle \int_{-\infty}^\infty \frac{\ud \w}{2\pi} \frac{\ex^{-\im \w t}}{(\ex^{\beta(\w-\mu)}+1)(\w-\eps_k)}.
\end{align}
Due to the exponential factor in the numerator, we close the integration contour in the lower-half plane. Since $\eps_k$ is an eigenvalue of $h_{\text{HF}}^{\text{eq}}+\im\varGamma/2$, it is located on the upper-half plane ($\varGamma$ is a positive-definite matrix). Then, in the lower-half plane, only the residues at the Matsubara poles, $\w=\w_n=\im\pi(2n+1)/\beta + \mu$ (with $n$ integer), contribute to the integral. This consideration is very similar to Ref.~\onlinecite{Tuovinen2019NJP}, and also here it is possible to show that the result can be written in terms of the hypergeometric function~\cite{NIST:DLMF}
\begin{align}\label{eq:hypf}
& \int_{-\infty}^\infty \frac{\ud \w}{2\pi} \frac{\ex^{-\im \w t}}{(\ex^{\beta(\w-\mu)}+1)(\w-\eps_k)} \nonumber \\
= & \ \frac{\ex^{-\im (\mu-\im\pi/\beta) t}}{\im\beta(\eps_k-\mu)-\pi} \times \nonumber \\
& {}_2 F_1 \left[1,\frac{1}{2}-\frac{\im\beta(\eps_k-\mu)}{2\pi},\frac{3}{2}-\frac{\im\beta(\eps_k-\mu)}{2\pi},\ex^{-2\pi t/\beta}\right].
\end{align}
After this manipulation, Eq.~\eqref{eq:hypf} can simply be inserted back into Eq.~\eqref{eq:freqintegral-wba} with a suitable rotation of the left/right eigenvectors~\cite{Tuovinen2019NJP}.

While, in this approach, the restart protocol (GKBA$|$IC) is consistent with the adiabatic switching (GKBA$|$AS) only when the WBA is a good approximation, this is still a fairly practical way of computing the initial contacting collision integral in Eq.~\eqref{eq:freqintegral} because it is considerably faster than numerical integration and can be performed to arbitrary numerical precision~\cite{Michel2008}. In addition, the GKBA approach is expected to be accurate in this regime due to the choice of propagators at the level of WBA. We have checked that the transient features presented in Figs.~\ref{fig:cyclobutadiene} and~\ref{fig:graphene} are very well represented also when using Eq.~\eqref{eq:freqintegral-wba} with Eq.~\eqref{eq:hypf}.

%

\end{document}